\documentclass[conference]{IEEEtran}
\IEEEoverridecommandlockouts

\usepackage{graphicx}
\usepackage{url}
\usepackage{booktabs}
\usepackage{colortbl}
\usepackage{mathtools}
\usepackage{multirow}
\usepackage{hyperref}
\usepackage{amsthm}

\newtheorem{example}{Example}

\begin{document}

\title{Bottom-Up Generation of Verilog Designs for Testing EDA Tools}

\author{\IEEEauthorblockN{Jo\~{a}o Victor Amorim Vieira}
\IEEEauthorblockA{\textit{UFMG}\\
Belo Horizonte, Brazil \\
joao.amorim@dcc.ufmg.br}
\and
\IEEEauthorblockN{Luiza de Melo Gomes}
\IEEEauthorblockA{\textit{UFMG}\\
Belo Horizonte, Brazil \\
luizademelo@dcc.ufmg.br}
\and
\IEEEauthorblockN{Rafael Sumitani}
\IEEEauthorblockA{\textit{Cadence Design Systems}\\
Belo Horizonte, Brazil \\
srafael@cadence.com}
\and
\IEEEauthorblockN{Raissa Maciel}
\IEEEauthorblockA{\textit{Cadence Design Systems}\\
Belo Horizonte, Brazil \\
raissa@cadence.com}
\and
\IEEEauthorblockN{Augusto Mafra}
\IEEEauthorblockA{\textit{Cadence Design Systems}\\
Belo Horizonte, Brazil \\
augusto@cadence.com}
\and
\IEEEauthorblockN{Mirlaine Crepalde}
\IEEEauthorblockA{\textit{Cadence Design Systems}\\
Belo Horizonte, Brazil \\
mirlaine@cadence.com}
\and
\IEEEauthorblockN{Fernando Magno Quintão Pereira}
\IEEEauthorblockA{\textit{UFMG}\\
Belo Horizonte, Brazil \\
fernando@dcc.ufmg.br}
}

\maketitle

\begin{abstract}
Testing Electronic Design Automation (EDA) tools rely on benchmarks—designs written in Hardware Description Languages (HDLs) such as Verilog, SystemVerilog, or VHDL.
Although collections of benchmarks for these languages exist, they are typically limited in size.
This scarcity has recently drawn more attention due to the increasing need for training large language models in this domain.
To deal with such limitation, this paper presents a methodology and a corresponding tool for generating realistic Verilog designs.
The tool, ChiGen, was originally developed to test the Jasper\textsuperscript{\tiny\textregistered} Formal Verification Platform, a product by Cadence Design Systems.
Now, released as open-source software, ChiGen has been able to identify zero-day bugs in a range of tools, including Verible, Verilator, and Yosys.
This paper outlines the principles behind ChiGen's design, focusing on three aspects of it: (i) generation guided by probabilistic grammars, (ii) type inference via the Hindley-Milner algorithm, and (iii) code injection enabled by data-flow analysis.
Once deployed on standard hardware, ChiGen outperforms existing Verilog Fuzzers such as Verismith, TransFuzz, and VlogHammer regarding structural diversity, code coverage, and bug-finding ability.
\end{abstract}


\section{Introduction}
\label{sec_intro}

Fuzzing~\cite{Takanen18} is an automated testing technique that generates random---often unexpected---inputs to a program to discover bugs, vulnerabilities, or unexpected behaviors.
Many EDA (Electronic Design Automation) tools such as \textsc{Yosys}, \textsc{Verilator}, \textsc{ModelSim}\textsuperscript{\tiny\textregistered}, \textsc{Xcelium}\textsuperscript{\tiny TM}, \textsc{Jasper}\textsuperscript{\tiny\textregistered}, and Design Compiler\textsuperscript{\tiny\textregistered} can benefit from a fuzzer that generates Verilog designs automatically.
This benefit comes in the form of early bug discovery, performance optimizations, compliance checking, and general validation.

There exist open-source Verilog fuzzers, such as VlogHammer~\cite{Wolf24}, Verismith~\cite{Herklotz20}, and TransFuzz~\cite{Solt25}. These tools operate in a top-down fashion: starting with a minimal core of valid Verilog syntax and expanding it through various techniques, always ensuring the generation of semantically valid designs. However, our experience using these tools to test the Jasper Verification Platform suggests that the requirement to produce only valid Verilog designs limits the diversity of their test cases. As outlined in Section~\ref{sec_ovf}, semantically invalid Verilog specifications can be equally effective as valid ones in uncovering issues in EDA tools. Additionally, these tools often produce syntactic constructs that are very different from those found in human-written code. For instance, as detailed in Section~\ref{sub_diversity}, these tools cover fewer than 40\% of the production rules in a grammar of Verilog-2005 (IEEE 1364-2005)~\cite{IEEE06}. \\

\textbf{Contributions of This Work:}  
This paper presents ChiGen, a ``bottom-up'' fuzzer designed to test the Jasper Formal Verification Platform from Cadence Design Systems.  
In October 2024, ChiGen was released as an open-source tool.
Since then, it has received contributions from academics and engineers, evolving into an effective Verilog fuzzer.  
Unlike state-of-the-art Verilog fuzzers, ChiGen generates designs in a bottom-up fashion.
It first produces a syntactically valid design with placeholders for user-defined symbols, such as variables, modules, and functions.
These placeholders are then replaced through multiple inference steps, transforming the skeleton into a valid Verilog design.  

As described in Section~\ref{sec_sol}, ChiGen operates in four stages:  
First, it generates the skeleton of a Verilog specification using a probabilistic grammar. The probabilities of production rules were trained over the benchmark suite detailed in Section~\ref{sec_chiBench}.
Second, ChiGen replaces mock identifiers with names that respect scoping rules.  
Third, it applies the Hindley-Milner type inference algorithm~\cite{Sulzmann00}, commonly used in functional programming, to infer the types of variables.
Finally, it employs a technique recently proposed by Li \emph{et al.}~\cite{Li24} to combine Verilog modules, functions, and references, achieving any predefined number of tokens.  

As discussed in Section~\ref{sec_eval}, ChiGen surpasses top-down fuzzers such as Verismith, VlogHammer, and TransFuzzer in code coverage, structural diversity, and bug-finding effectiveness. Since its release, it has uncovered issues, such as those described in Section~\ref{sec_ovf}, in open-source tools such as Yosys, Verilator, Icarus, and Verible. These successes stem from the following contributions:  
\begin{itemize}
\item \textbf{Real-World Training Set:} ChiGen emulates the syntax of real Verilog designs. 
To train it, we curated a dataset of 50,000 designs mined from open-source repositories with permissible licenses.  
This benchmark suite, referred to as ChiBench, is a contribution in itself and has been used for tasks beyond training ChiGen, as discussed in Section~\ref{sec_chiBench}.

\item \textbf{Probabilistic Grammar:} As detailed in Section~\ref{sub_prob_grammar}, ChiGen can be trained on any number of Verilog examples without human intervention. The more designs it observes during training, the more realistic the designs that it generates.

\item \textbf{Inference Mechanism:} A trained instance of ChiGen produces syntactically valid Verilog skeletons, which are then refined through static analyses. These analyses ensure that variables are defined before use (Section~\ref{sub_scope}) and that all references adhere to declared types (Section~\ref{sub_type_infer}).  

\item \textbf{Code Injection:} ChiGen incrementally integrates generated components into more complex designs, linking modules via bindings, instantiations, function calls or hierarchical references (Section~\ref{sub_code_injection}). As new modules, functions, and references are generated, they become available for insertion into the ongoing design.  
\end{itemize}  
The ChiGen fuzzer and the ChiBench suite of Verilog designs are publicly available under the GPL 3.0 license and can be retrieved at \url{https://github.com/lac-dcc/chimera}.

\section{Overview}
\label{sec_ovf}

\begin{figure}[ht]
\includegraphics[width=\columnwidth]{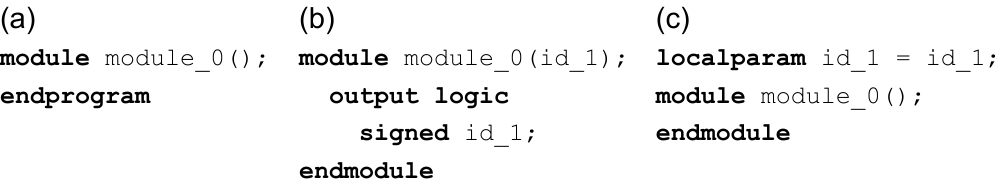}
\caption{(a-b) Designs that uncovered issues in \textsc{Verible}.
(c) Design that uncovered issues in \textsc{YoSys}.}
\label{fig_example_bugs}
\end{figure}

This section illustrates the usage of ChiGen with the three designs seen in Figure~\ref{fig_example_bugs}.
These files were automatically produced by ChiGen, using a probabilistic grammar with a context of length one (the notion of context depth shall be explained in Section~\ref{sub_prob_grammar}).
All these designs have uncovered an issue in some EDA tool.
All the issues were reported and acknowledged.
Throughout this paper, we shall say that a Verilog design is valid if it passes Jasper's analysis phase (e.g., it successfully go through the \texttt{analyze} command).
Thus, invalid designs either show incorrect syntax or fail the static semantic analysis. \\

\noindent
\textbf{Tool Accepts Invalid Syntax: }
The first code, in Figure~\ref{fig_example_bugs} (a) uncovered a bug in Verible's parser.
The keyword \texttt{endprogram} incorrectly matches the keyword \texttt{endmodule}.
However, even though the design is syntactically invalid, it was accepted by Verible's parser.
In this case, the expected behavior would be to report the syntactic error.

\noindent
\textbf{Tool Crashes on Valid Syntax: }
The code in Figure~\ref{fig_example_bugs} (b) is valid; however, the port \texttt{id\_1 }is declared as \texttt{output} \texttt{logic} \texttt{signed}.
This syntax is not typical in traditional Verilog but is acceptable in SystemVerilog.
Nevertheless, this specification causes a segmentation fault in Verible's parser. \\

\noindent
\textbf{Tool Crashes on Invalid Semantics: }
The design in Figure~\ref{fig_example_bugs} (c) is syntactically valid.
The syntax for declaring a \texttt{localparam} outside any module is acceptable.
However, the code is not semantically valid.
The issue lies in the \texttt{localparam} declaration: the line \texttt{localparam} \texttt{id\_1} = \texttt{id\_1;} attempts to assign the value of \texttt{id\_1} to itself, which creates a circular reference.
In Verilog, \texttt{localparam} must be initialized to a constant expression or a value known at compile time; hence, referring to itself in this way does not provide a valid initialization value.
This module causes infinite recursion in Yosys, eventually forcing a crash, as the tool runs out of stack memory.

In addition to the three examples described in this section, designs generated by ChiGen have discovered several other issues in popular EDA tools, including Icarus, Verilator, Verible's obfuscator, and Verible's formatter.
Some of these issues have led to non-trivial changes in these tools.
As an example, a ChiGen design provoked the addition of syntactic rules in Verible's parser to
mix anonymous and named instances.
Similarly, at least two interventions were recently added to Verilator (Release 5.030 2024-10-27) due to issues raised via ChiGen.
ChiGen has also been successful in discovering issues in proprietary tools.
However, in this paper, we will only discuss issues that have been publicly reported.

\section{Bottom-Up Fuzzing}
\label{sec_sol}

\begin{figure}[ht]
\includegraphics[width=\columnwidth]{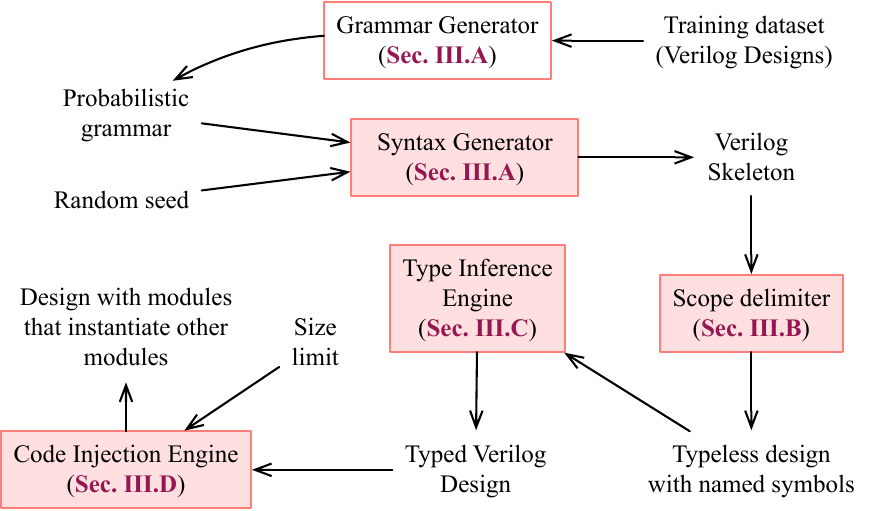}
\caption{Overview of ChiGen's modus operandi.}
\label{fig_ChiGenOvf}
\end{figure}

ChiGen works in four phases.
Figure~\ref{fig_ChiGenOvf} shows how these phases are related.
The rest of this section describes each of these steps.

\subsection{Syntax Generation via Probabilistic Grammars}
\label{sub_prob_grammar}

%A probabilistic context-free grammar (PCFG) is an extension of a traditional context-free grammar that assigns probabilities to each production rule. These probabilities represent the likelihood of each rule been taken; hence, the probabilities associated with all production rules for a given non-terminal must sum to 1.0. 

%\noindent \textbf{The Source Grammar} To parse Verilog designs, the Grammar Generator uses a modified version of Verible's Verilog grammar, which is implemented in Bison. We had to manually modify Verible's grammar, because this grammar was conceived to recognize a superset of valid Verilog designs. It was not conceived to generate only valid Verilog designs. Thus, without our modifications, the original grammar could generate invalid syntax, such as interposing ANSI and non-ANSI port declarations.

To produce a Verilog specification, ChiGen begins by generating a ``skeleton'' of the design: a structure that adheres to Verilog's syntactic rules. This skeleton is created using a Probabilistic Context-Free Grammar (PCFG), which assigns probabilities to sequences of production rules.
These production rules were taken from Verible's grammar, which parses the IEEE 1800-2017 standard.
As seen in Figure~\ref{fig_ChiGenOvf}, ChiGen is distributed with a grammar generator. This tool receives a training set (a collection of Verilog designs). It parses every design in the training set, recording the number of times each grammar production was activated during parsing as a YAML file. The grammar generator uses this log to build the probabilities associated with each production rule in the Verilog grammar.
The public distribution of ChiGen was trained with a collection of designs extracted from a benchmark suite called ChiBench.
The construction of ChiBench is the subject of Section~\ref{sec_chiBench}. \\

\noindent \textbf{Context-Sensitive Probabilities} In a traditional PCFG, each production rule's probability is independent of the others, making it ``memoryless'' (or Markovian).
Thus, the probability of applying a rule to a non-terminal depends only on the non-terminal itself, not on preceding or succeeding rules. However, context-sensitive probabilistic models allow for conditional dependencies across rule applications, where the probability of a rule can depend on previously chosen rules, leading to ``sequence-aware'' probabilities.
ChiGen enables conditional dependencies between rule applications, allowing the probability of a given rule to depend on previously selected $K$ rules (a K-gram), resulting in ``sequence-aware'' probabilities.
We limit the probability context $K$ -- the chain of production rules associated with a probability -- to six productions, as each additional context introduces a potentially exponential increase in the table of probabilities.
To construct the PCFG, ChiGen parses a training set of Verilog designs.
It parses each file in this set, recording how often each sequence of productions is used during parsing.
Example~\ref{ex_probabilisticGrammar} shows instances of probabilistic grammars.

\begin{figure}[ht]
\includegraphics[width=\columnwidth]{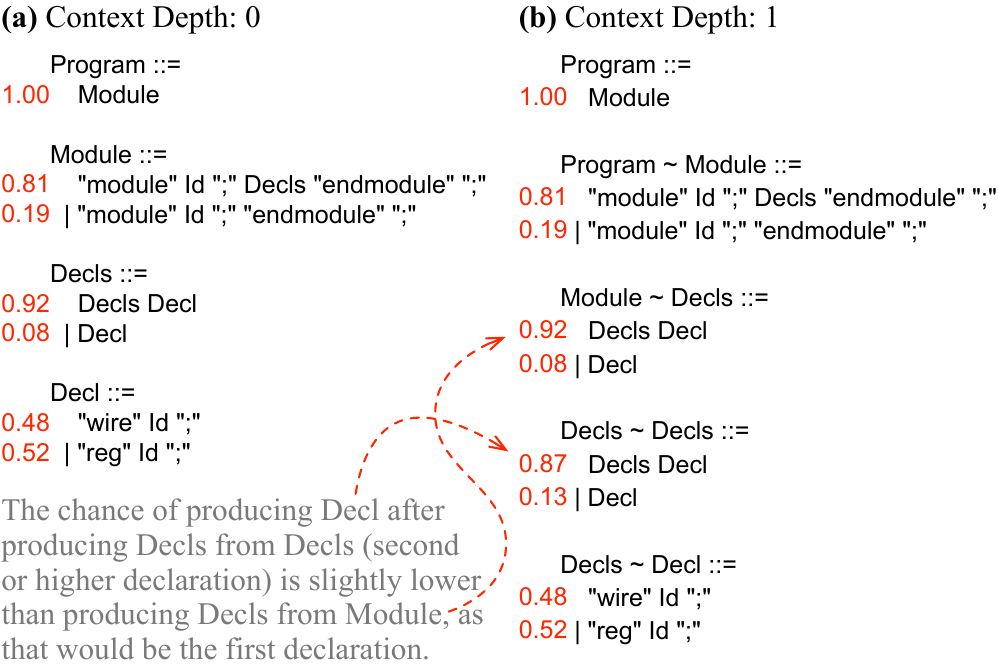}
\caption{Two probabilistic grammars. They recognize the same language, albeit with two different contexts of probabilities.}
\label{fig_probabilisticGrammar}
\end{figure}

\begin{example}
\label{ex_probabilisticGrammar}
Figure~\ref{fig_probabilisticGrammar} shows two examples of PCFGs.
The example in Figure~\ref{fig_probabilisticGrammar} (a) does not take context into consideration.
The example in Figure~\ref{fig_probabilisticGrammar} (b) considers contexts of depth one; that is, it can ``remember'' the rule that led to the production of the current nonterminal that must be expanded.
As an illustration, the chance of adding a new element to a list of declarations decreases if we know that this list already has one element, as very long chains of declarations are uncommon.
\end{example}

Each production rule is activated according to its probabilities.
The starting symbol of the grammar has probability 1.0; hence, syntax generation always starts with a non-null design.
If a rule $A ::= B C$ is activated, then it creates two new nonterminals, which will, in turn, also be activated.
Each of these nonterminals might be the left-hand side of multiple productions.
The choice of which production is activated depends on the probabilities associated with them.
This process terminates, as eventually terminals, or the empty string, are produced.
At the end of syntax generation, we obtain a skeleton of a Verilog design, as Example~\ref{ex_VerilogSkeleton} illustrates.

\begin{figure}[ht]
\includegraphics[width=\columnwidth]{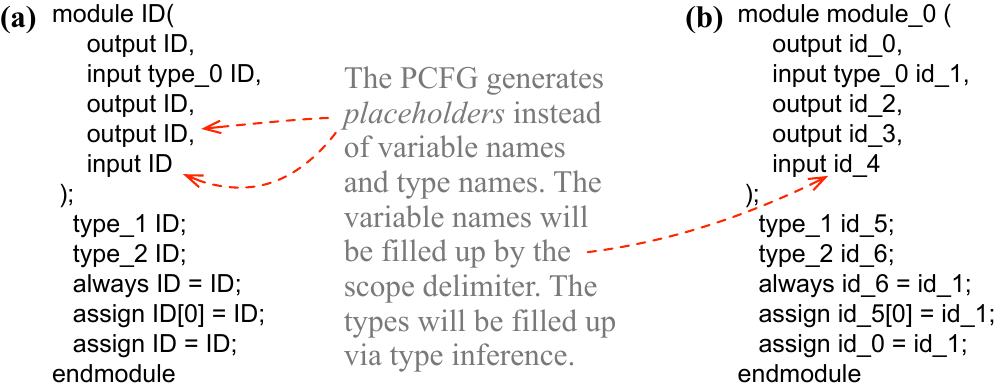}
\caption{Skeleton generation and symbol renaming.}
\label{fig_VerilogSkeleton}
\end{figure}

\begin{example}
\label{ex_VerilogSkeleton}
Figure~\ref{fig_VerilogSkeleton} (a) shows an example of a design that is produced by exercising the probabilistic grammar.
Notice that this design is not valid, among other things, because every symbol is referred to as a placeholder.
\end{example}

\subsection{Variable Renaming and Scope Creation}
\label{sub_scope}

The design in Figure~\ref{fig_VerilogSkeleton} (a) contains only placeholders where symbol names are expected.
In the next state of code generation, a ``scope delimiter'' replaces these placeholders with variable names.
Renaming uses a set of ``in-scope'' variables and follows three rules:
(i) The declaration of a placeholder is renamed with a new symbol $s$, and $s$ is inserted into the set of in-scope elements associated with the current scope.
(ii) Uses of a placeholder are randomly replaced with symbols, respecting only their direction. Input identifiers should never appear on the left-hand side of assignments, and the contrary holds for output signals.
(iii) Once the scope delimiter leaves a scope region, it removes from the set of in-scope elements the variables declared within that region.
Example~\ref{ex_VerilogSkeleton_b} shows how these rules are applied in practice.

\begin{example}
\label{ex_VerilogSkeleton_b}
Figure~\ref{fig_VerilogSkeleton} (b) shows the effect of applying renaming on the skeleton earlier discussed in Example~\ref{ex_VerilogSkeleton}.
Variable usages, initially represented by generic ID names, are replaced with any of the user-defined symbols \texttt{ID\_0} to \texttt{ID\_6}, which are in-scope.
\end{example}

\textbf{Dealing with Instantiable Namespaces.}
An \textit{instantiable namespace} is a programming construct that defines a scope containing variables, functions, or other elements, where multiple independent instances of this scope can be created.
Unlike noninstantiable namespaces, which provide a global or hierarchical organization of names (e.g., \texttt{namespace} in C++ or \texttt{package} in Java), instantiable namespaces allow for multiple copies, each maintaining its own state.
Examples include \texttt{struct} or \texttt{union} in C, and \texttt{class} in Python.
Verilog provides one form of instantiable namespace in the \texttt{module} construct.
SystemVerilog supports, additionally, interfaces, structs, unions, and classes.
ChiGen is currently able to produce modules, unions, and structs (packed or unpacked).
The presence of instantiable namespaces has an impact on the implementation of the scope delimiter, which must keep a table with all the namespaces instantiated in the current scope.
Example~\ref{ex_namespace} illustrates this feature.

\begin{figure}[ht]
\includegraphics[width=\columnwidth]{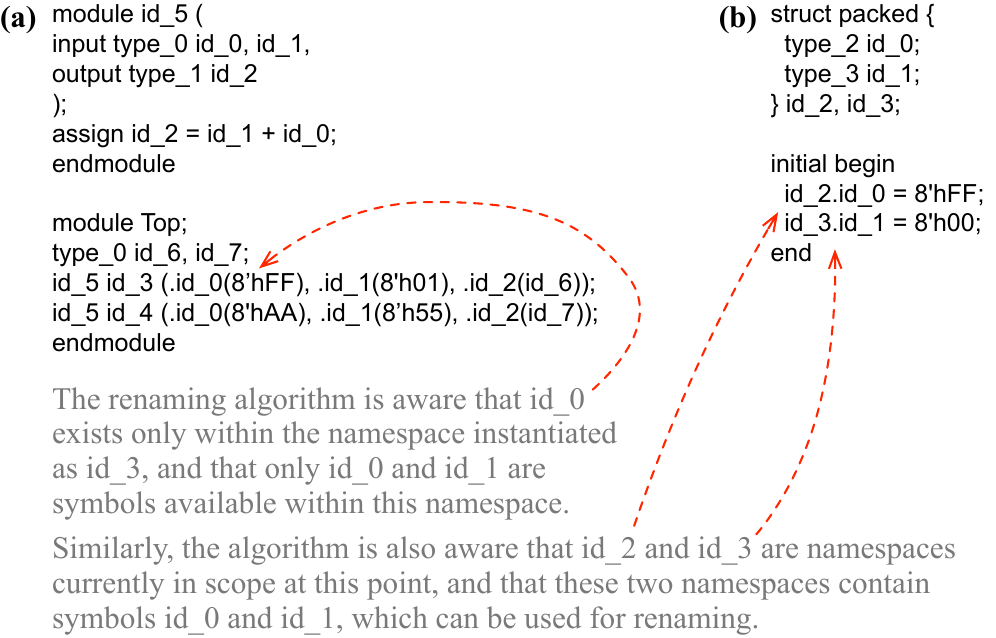}
\caption{Example of synthetic designs with instantiable namespaces.}
\label{fig_namespaces}
\end{figure}

\begin{example}
\label{ex_namespace}
Figure~\ref{fig_namespaces} (a) shows a synthetic design with references to names defined within a module.
Figure~\ref{fig_namespaces} (b) shows a design with names defined within a SystemVerilog \texttt{struct}.
Both constructions are supported by ChiGen.
\end{example}

\subsection{Type Inference via Unification}
\label{sub_type_infer}

The scope delimiter in Section~\ref{sub_scope} assigns names to variables, but their types remain undefined.
In the subsequent phase of code generation, a ``type inference engine'' deduces these types. This type inference process follows the well-known Hindley-Milner algorithm, which is widely used in languages such as SML/NJ, Haskell, and Rust. However, we adopt the two-stage formulation proposed by Sulzmann~\cite{Sulzmann00}: first, we generate constraints; then, we solve these constraints through unification.
Each constraint consists of a pair $(t_0, t_1)$, indicating that the terms $t_0$ and $t_1$ must share the same type. These terms may represent primitive types or open type variables (such as \texttt{type\_1} in Figure~\ref{fig_VerilogSkeleton}).
Example~\ref{ex_typeInference} summarizes this process, while the rest of this section provides
details on each one of its two phases.

\begin{figure}[ht]
\includegraphics[width=\columnwidth]{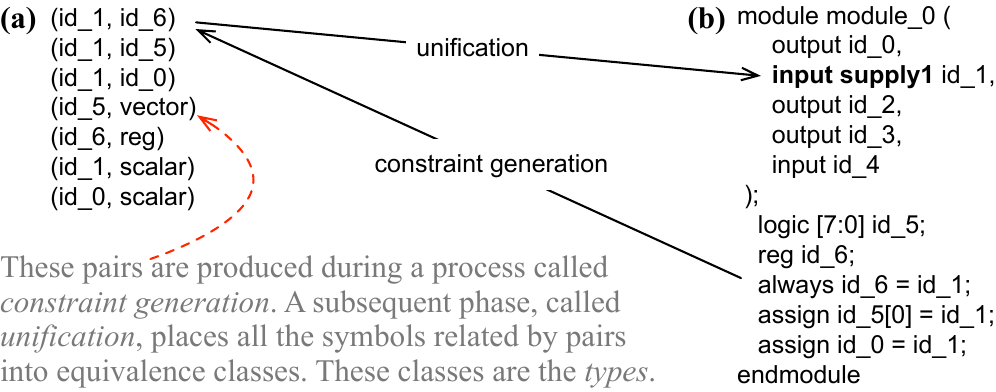}
\caption{Hindley-Milner Type Inference.}
\label{fig_typeInference}
\end{figure}

\begin{example}
\label{ex_typeInference}
Figure~\ref{fig_typeInference} (a) shows the seven pairs of constraints generated for the design in Figure~\ref{fig_VerilogSkeleton} (b). These pairs are produced by visiting the abstract syntax tree that describes the skeleton code. For instance, the pair $(\mathtt{id\_1}, \mathtt{id\_6})$ is produced because of the assignment \texttt{always id\_6 = id\_1} present in the skeleton. This pair indicates that the type of these two identifiers must be the same. The result of unifying all the pairs appears in Figure~\ref{fig_typeInference} (b), where the type placeholders have been replaced with actual type names in this updated version of our running example.
\end{example}

\textbf{Constraint Generation.}
The process of constraint generation in Hindley-Milner type inference follows a visitor-like traversal of the abstract syntax tree (AST). As the visitor encounters each node, it introduces a set of type variables representing the types of identifiers and expressions at that node. Additionally, it generates a set of constraints, each of which is a pair expressing a type equivalence or compatibility requirement between two symbols—these symbols can be either type variables or user-defined identifiers. The constraints ensure that operations receive operands of appropriate types and that results are correctly propagated throughout the AST.
Example~\ref{ex_constraintGen} explains this step.

\begin{example}
\label{ex_constraintGen}
When visiting an abstract syntax tree (AST) node, such as \texttt{assign id\_6 = id\_1 + id\_2;}, the constraint generation engine proceeds as follows:

\begin{enumerate}
    \item Create fresh type variables for each identifier in the assignment: \( t_{id6} \) for \texttt{id\_6}, \( t_{id1} \) for \texttt{id\_1}, and \( t_{id2} \) for \texttt{id\_2}

    \item Create three constraints to associate each identifier with its type variable:
    \(
    (id_6, t_{id6}), (id_1, t_{id1}), (id_2, t_{id2})
    \)

    \item Create a constraint to enforce operand type compatibility:
    \(
    (t_{id1}, t_{id2})
    \)

    \item Define two conditional constraints for result type inference:
    \(
    (t_{id1}, \text{bitvector}(N)) \Rightarrow (t_{id6}, \text{bitvector}(N+1)),
    (t_{id2}, \text{bitvector}(N)) \Rightarrow (t_{id6}, \text{bitvector}(N+1))
    \)

\end{enumerate}
These ensure that:
\begin{itemize}
    \item The operands have the same type.
    \item The result follows Verilog’s bit-width extension rules.
\end{itemize}
\end{example}

\textbf{Unification.}
Constraints are solved via a process called ``{\it Unification}''.
Unification finds a substitution of type variables by actual types that makes all the equalities hold.
The final product of unification is a table -- also known as an ``{\it environment}'' --
that associates type variables with actual types.
To build this table, the unification engine iterates through the constraints, applying known type assignments and propagating these assignments throughout the system, as Example~\ref{ex_unification} shows.
If a conflict arises -- such as trying to unify \texttt{bitvector(8)} with \texttt{bitvector(16)} -- then the unification fails, indicating a type error.
In this event, ChiGen discards the current Verilog skeleton.

\begin{example}
\label{ex_unification}
Continuing with Example~\ref{ex_constraintGen}, the type inference engine must solve the constraints that were produced for \texttt{assign id\_6 = id\_1 + id\_2;}.
Assume that throughout the type resolution process, the unification engine already has in the unification table the assumption that $t_{id1} = \text{bitvector}(8)$.
This assumption, plus the constraint $(t_{id1}, t_{id2})$ gives us that
$t_{id2} = \text{bitvector}(8)$.
Thus, substituting into the first constraint $(id_2, t_{id2})$, we conclude that
$t_{id2} = \text{bitvector}(8)$.
Then, applying the second constraint
$(t_{id2}, \text{bitvector}(N)) \Rightarrow (t_{id6}, \text{bitvector}(N+1))$, we conclude that $t_{id6} = \text{bitvector}(9)$.
\end{example}

If insufficient constraints are available to determine the type of an identifier, then we use \texttt{wire} as the default type according to the IEEE 1800-2017 standard rule for nets declared without an explicit type.
Nevertheless, even with such an expedient, some Verilog skeletons cannot undergo type inference successfully.
Type inference may fail if constraints require the unification of two incompatible primitive types.
When type inference fails, the skeleton is discarded, and the random seed that produced it is used as input to generate a new seed. The failure rate is influenced by the probabilistic grammar used.
In the experiments described in Section~\ref{sec_eval}, the chosen grammar yields a success rate of 50\% with a probabilistic context of length one and 75\% with a probabilistic context of length three.

\subsection{Code Expansion via Code Injection}
\label{sub_code_injection}

To control the size of Verilog designs generated by ChiGen, we use a technique called {\it code injection}, following the approach introduced by Li {\it et al}~\cite{Li24} in 2024.
Code injection involves combining multiple designs to create a new, syntactically and semantically valid design.
In this case, a {\it caller} block invokes a {\it callee} unit using 
some syntax available in the target programming language.
Currently, ChiGen supports the following kinds of code injection:
\begin{itemize}
\item \textbf{Module instantiation:} a caller module $M_{caller}$ invokes a callee module $M_{callee}$.
\item \textbf{Function invocation:} a caller module $M_{caller}$ invokes a function declared inside of it.
\item \textbf{Hierarchical references:} a caller module $M_{caller}$ refers to a symbol $R$ declared within a callee module $M_{callee}$ or vice versa.
\end{itemize}
Example~\ref{ex_codeInjection} shows the last two forms of injections.
Example~\ref{ex_moduleInjection}, at the end of this section, illustrates the first.

\begin{figure}[ht]
\includegraphics[width=\columnwidth]{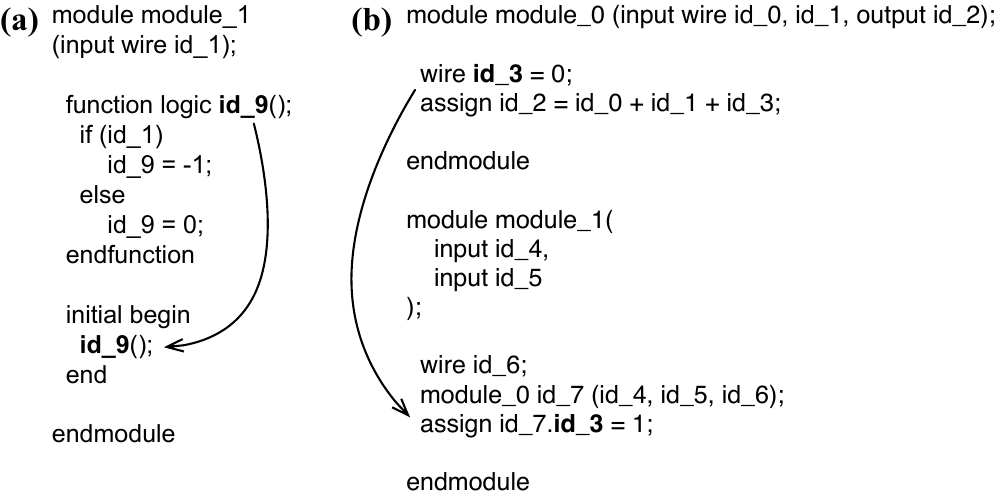}
\caption{(a) Injection of function call.
(b) Injection of hierarchical reference.}
\label{fig_codeInjection}
\end{figure}

\begin{example}
\label{ex_codeInjection}
Figure~\ref{fig_codeInjection} illustrates two forms of code injection that can be present in designs produced by ChiGen.
In part (a), a function \texttt{id\_9} is called within an initial block, demonstrating procedural interaction where the function's logic depends on input \texttt{id\_1}.
In part (b), module \texttt{module\_1} instantiates \texttt{module\_0} and directly manipulates its internal wire \texttt{id\_3} via hierarchical assignment (assign \texttt{id\_7.id\_3 = 1}), showcasing structural interaction between modules.
\end{example}

\begin{figure}[ht]
\includegraphics[width=\columnwidth]{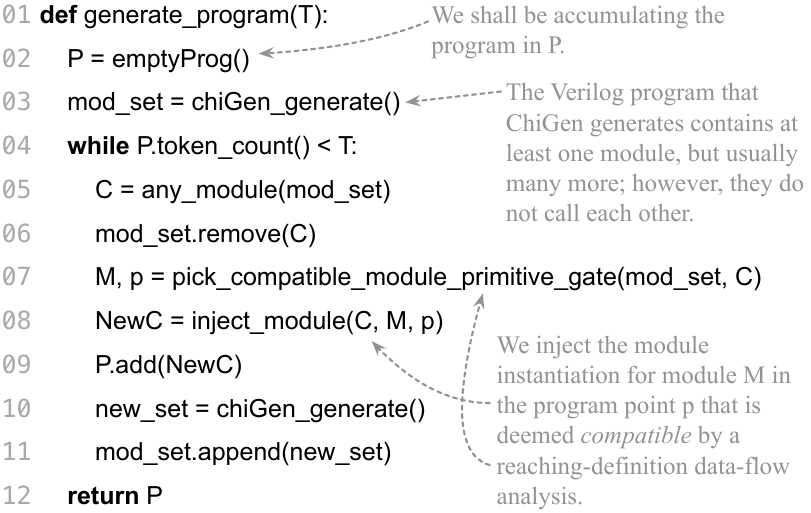}
\caption{The algorithm that implements module injection: it injects an instantiation of .}
\label{fig_inject_module}
\end{figure}

Figure~\ref{fig_inject_module} describes our algorithm that implements module injection (injection of function calls and hierarchical references follow a similar approach).
In Li {\it et al.}'s method, a new program $P$ is built by combining two existing programs, $P_0$ and $P_1$, from a real-world project.
In contrast, we perform module injection interactively: as shown in Figure~\ref{fig_inject_module}, we start with an empty design $P$ and continue adding new modules to it, via the \texttt{chiGen\_generate} routine, until the design reaches a preset token count, $T$.
Notice that the \texttt{inject\_module} function inserts new syntax into an existing module: this new syntax implements the instantiation rule that connects two program units, like modules, functions, or hierarchical references.

\textbf{Reaching Definition Analysis}
Following Li {\it et al.}'s approach, we use the reaching definition data-flow analysis to determine where and how to inject modules into the accumulated program $P$. Reaching definition associates each program point $p \in P$ with the set of variables that reach $p$. A variable $v$ reaches a program point $p$ if the program $P$ contains a path from the definition of $v$ until the point $p$, and $v$ is not redefined along this path.
We can inject a module $M$ at a program point $p \in P$ if, and only if, for each input (respectively, output) parameter $a$ of $M$, there is an input (respectively, output) variable $v$ of equivalent type reaching $p$. When there are multiple such variables, we pick any of them randomly. Our module injection procedure prevents cycles in the final call graph by removing a module from the list of available modules once it is injected, as shown on Line 06 of Figure~\ref{fig_inject_module}.
One last observation about module injection refers to the fact that we can, at any given time, pick a Verilog primitive gate (\texttt{or}, \texttt{and}, \texttt{xor}, etc) instead of a module in \texttt{mod\_set} to inject.
The function \texttt{pick\_compatible\_module\_primitive\_gate} in Line 07 chooses a primitive gate or a module based on the probabilities found in the training set used in Section~\ref{sub_prob_grammar}.
Notice that primitive gates are not part of \texttt{mod\_set}, since they are defined in the Verilog language; hence, they are never removed from the pool of modules available for injection.

\begin{figure}[ht]
\includegraphics[width=\columnwidth]{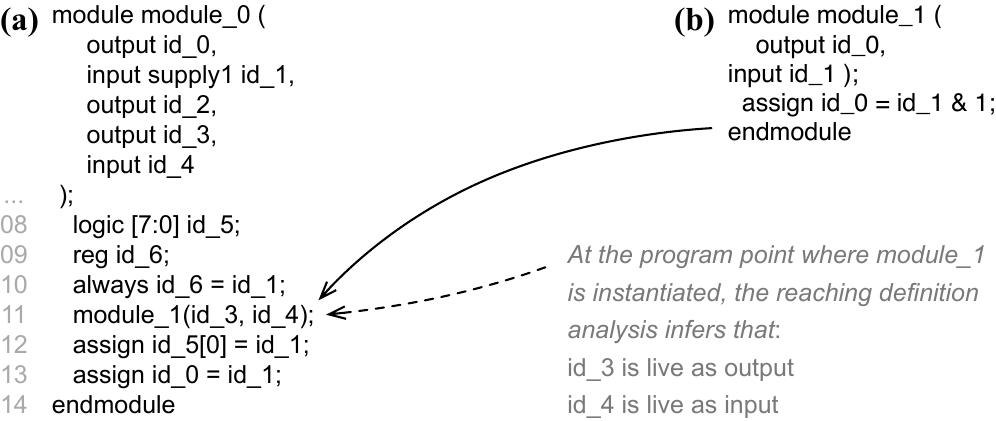}
\caption{Example of a design after module injection.}
\label{fig_moduleInjection}
\end{figure}

\begin{example}
\label{ex_moduleInjection}
Figure~\ref{fig_moduleInjection} shows how the reaching-definition analysis enables
code injection.
Variables \texttt{id\_3} and \texttt{id\_4} reach Line 11 in Figure~\ref{fig_moduleInjection} (a).
These variables are compatible with the signature of \texttt{module\_1}, which is part of the pool of modules available for injection.
Hence, an instantiation of \texttt{module\_1} is inserted at Line 11 of \texttt{module\_0}.
\end{example}

\section{ChiBench: The Training Set}
\label{sec_chiBench}

The probabilistic grammar used in the open-source distribution of ChiGen (see Section~\ref{sub_prob_grammar}) was trained on a dataset of 10,000 designs.  
These designs were selected from ChiBench, a larger collection of 50,000 designs curated specifically for training ChiGen.  
To extract this subset, we sorted the ChiBench designs by size, measured in terms of the number of tokens.  
From this ordered sequence, we selected the 5,000 designs immediately below the median and the 5,000 designs immediately above it.  
This approach helps avoid outliers -- designs that are either too small or too large -- while keeping training times manageable.
This section details the methodology used to construct the ChiBench collection.

In order to build ChiBench, we have mined designs from open-source GitHub repositories, using
GitHub's REST API. We
use GitHub's API to build a list of candidate Verilog repositories.
This list is sorted by popularity (measured as the number of stargazers).
We remove from the candidate list repositories that are not available for public usage, due to the lack of a license.
Thus, for each repository $R$ in the sorted list, we have implemented a Python script that proceeds as follows:

\begin{enumerate}
    \item Clone $R$ and locally copy all its \texttt{.v} files;
    \item Assigns a unique name to each \texttt{.v} file, based on its repository and its local path;
    \item Remove any special characters from the file's name to avoid encoding issues.
\end{enumerate}
We repeat the above sequence of steps for all the repositories in the base list,
until a predefined number of files is reached.

\subsection{Curating the Data}
\label{sub:curate}

After we have copied the necessary number of Verilog files from GitHub, we proceed to select valid designs.
To this effect, we only keep files that are syntactically and semantically valid.
Thus, this process involves passing the files through two sieves.
The first sieve, the syntax analysis, happens via the Verible syntactic analyzer.
At this stage, if Verible's parser cannot build an abstract syntax tree for a file, we discard it.
Example~\ref{ex_syntactically_invalid_v} illustrates one such situation.

\begin{figure}[ht]
\centering
\includegraphics[width=\columnwidth]{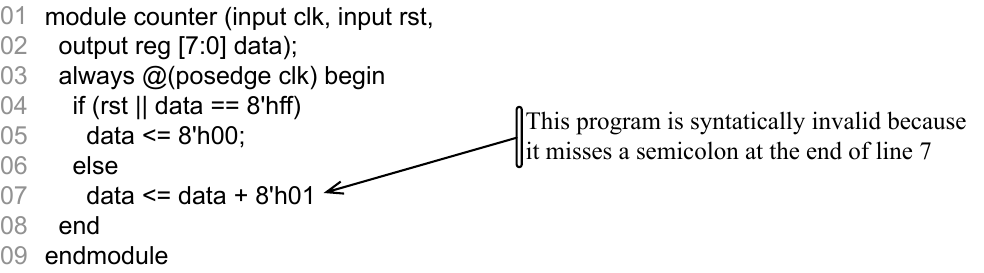}
\caption{Specification filtered out by syntactic verification.}
\label{fig_syntactically_invalid_v}
\end{figure}

\begin{example}
\label{ex_syntactically_invalid_v}
The design in Figure~\ref{fig_syntactically_invalid_v}, which specifies an 8-bit counter, will be filtered out by the syntactic filter.
It contains a missing semicolon at Line 7.
Such syntactically invalid files might occur in the mining process, as the repositories contain, for instance, files that are still under development.
\end{example}

Once we remove any syntactically invalid designs, we use Jasper's HDL semantic analyzer to filter out any semantically invalid designs.
Notice that Jasper's HDL analyzer also rejects invalid syntax.
However, Jasper's HDL analyzer is more computationally expensive than Verible's because it also considers semantic analysis, being more restricted to the Verilog language standard. Consequently, to reduce the number of designs sent for semantic analysis, we chose to filter out syntactically invalid designs before using Jasper.
Example~\ref{ex_semantically_invalid_v} better explains the semantic analysis's role.

\begin{figure}[ht]
\centering
\includegraphics[width=\columnwidth]{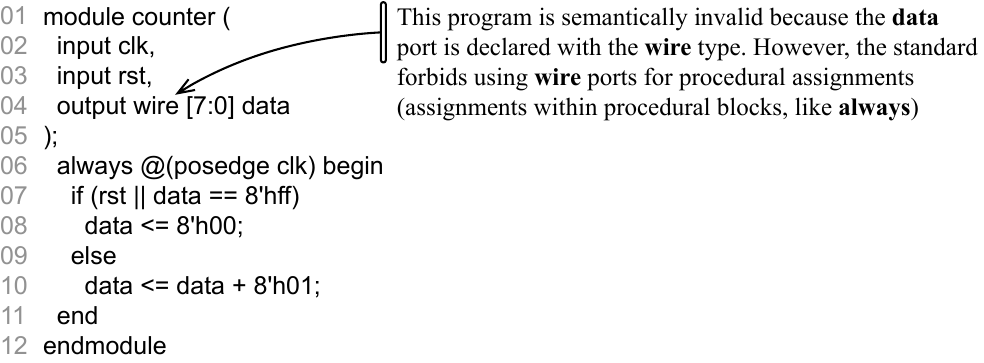}
\caption{Verilog specification that fails the semantic test.}
\label{fig_semantically_invalid_v}
\end{figure}

\begin{example}
\label{ex_semantically_invalid_v}
Figure~\ref{fig_semantically_invalid_v} shows an example of a design that fails the semantic sieve due to a type inconsistency.
In this case, the IEEE standard forbids the declaration of data ports with the wire type.
Thus, this design is invalid because it is trying to assign a value to \texttt{data} inside an \texttt{always} block, but \texttt{data} is declared as an output wire. In Verilog, wires cannot be assigned inside an \texttt{always} block; only \texttt{reg} or \texttt{logic} types can be written values in procedural contexts.
\end{example}

\section{Evaluation}
\label{sec_eval}

This section evaluates the following research questions:

\begin{description}
\item[RQ1:] How diverse are the designs that ChiGen generates?
\item[RQ2:] How do ChiGen's designs compare with real-world codes in terms of the coverage that they enable in typical EDA tools?
\item[RQ3:] Which kinds of bugs can be uncovered via ChiGen-enabled fuzzing?
\item[RQ4:] What is ChiGen's throughput, measured in terms of Verilog designs produced per second?
\item[RQ5:] What is the effectiveness of the different techniques listed in Section~\ref{sec_sol} to increase the percentage and size of valid Verilog designs?
\end{description}

\paragraph{Baselines}
We have used 10,000 ChiBench designs (Section~\ref{sec_chiBench}) to train ChiGen's probabilistic grammar.
We compare ChiBench with Verismith~\cite{Herklotz20}, TransFuzz~\cite{Solt25} and VlogHammer~\cite{Wolf24}, which are other fuzzer collections.
VlogHammer always generates the same 3,000 designs.

\subsection{RQ1 -- Diversity}
\label{sub_diversity}

%The more diverse the designs generated by ChiGen, the greater the likelihood of uncovering different bugs in EDA tools.
This section evaluates the {\it syntactical diversity} of ChiGen-generated designs. In Section~\ref{sub_coverage}, we will examine—indirectly—their semantic diversity by assessing the coverage they enable when used as input files for EDA tools.
We measure syntactical diversity as the number of unique production rules in the Verible grammar required to parse these files.

\begin{figure}[ht]
\includegraphics[width=\columnwidth]{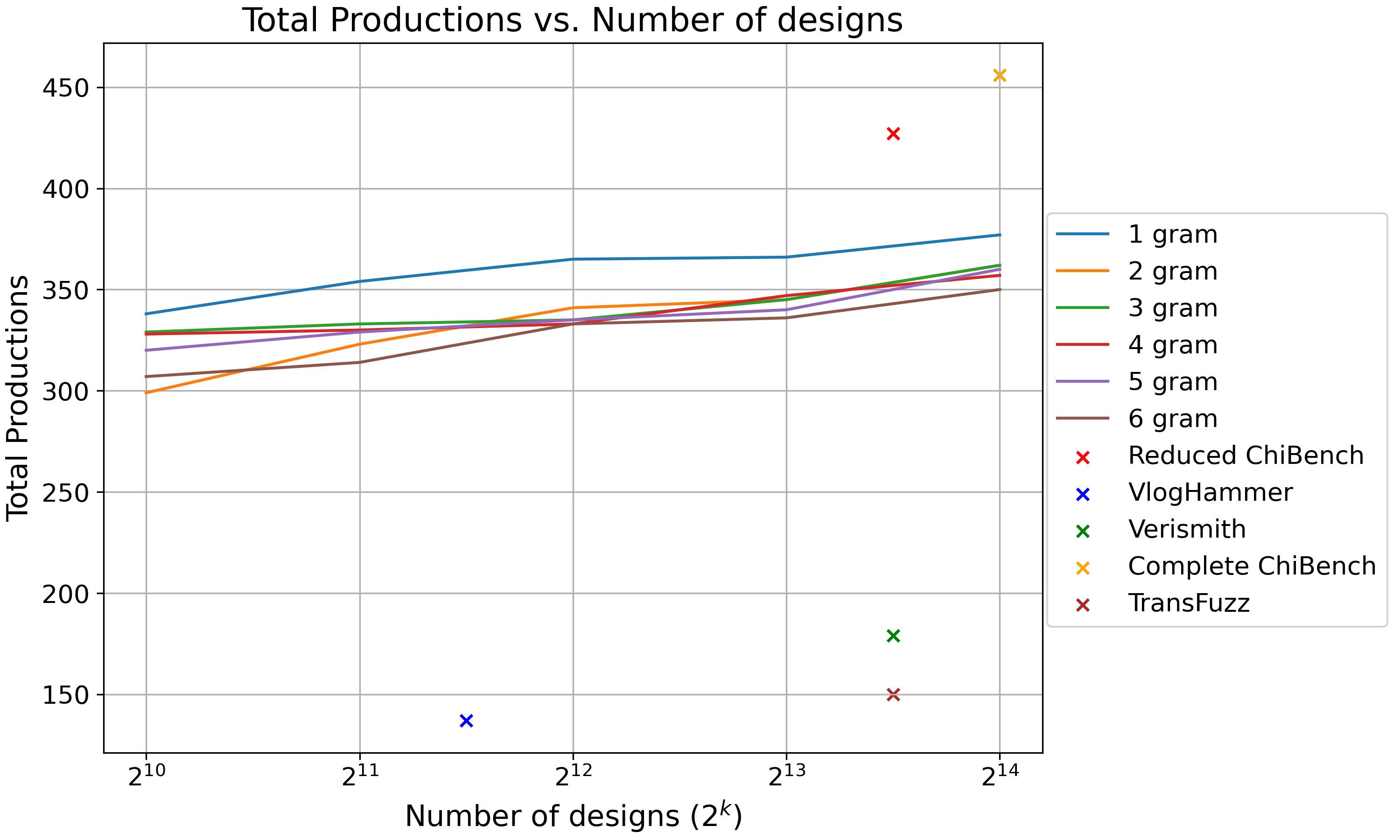}
\caption{Syntactical diversity of ChiGen designs, measured as the number of unique production rules in the Verilog grammar exercised when parsing a population of generated files.}
\label{fig_diversity}
\end{figure}

\paragraph{Discussion}
Figure~\ref{fig_diversity} shows the syntactical diversity across populations of various sizes of ChiGen designs. The figure counts unique production rules, meaning that multiple occurrences of the same rule (e.g., Decl ::= ``\texttt{wire}'' Id ``\texttt{;}'') within a population are counted only once. We observe that as the number of generated designs grows, the number of unique production rules used also increases, approaching the number found in ChiBench, our ground truth. In contrast, VlogHammer, Verismith, and TransFuzz exercise significantly fewer production rules.

In populations of 10,000 designs, ChiBench exercises 406 unique production rules, increasing to 456 in the complete dataset. Among the fuzzers, Verismith exercises 179 and TransFuzz, 151. VlogHammer -- which is limited to 3,000 designs -- uses 137 unique productions. ChiGen's performance varies slightly with the size of the probabilistic context $K$: for $K=1,2,3,4,5,6$, we observe 377, 362, 362, 357, 360, and 350 unique productions exercised, respectively.

\begin{figure}[ht]
\includegraphics[width=\columnwidth]{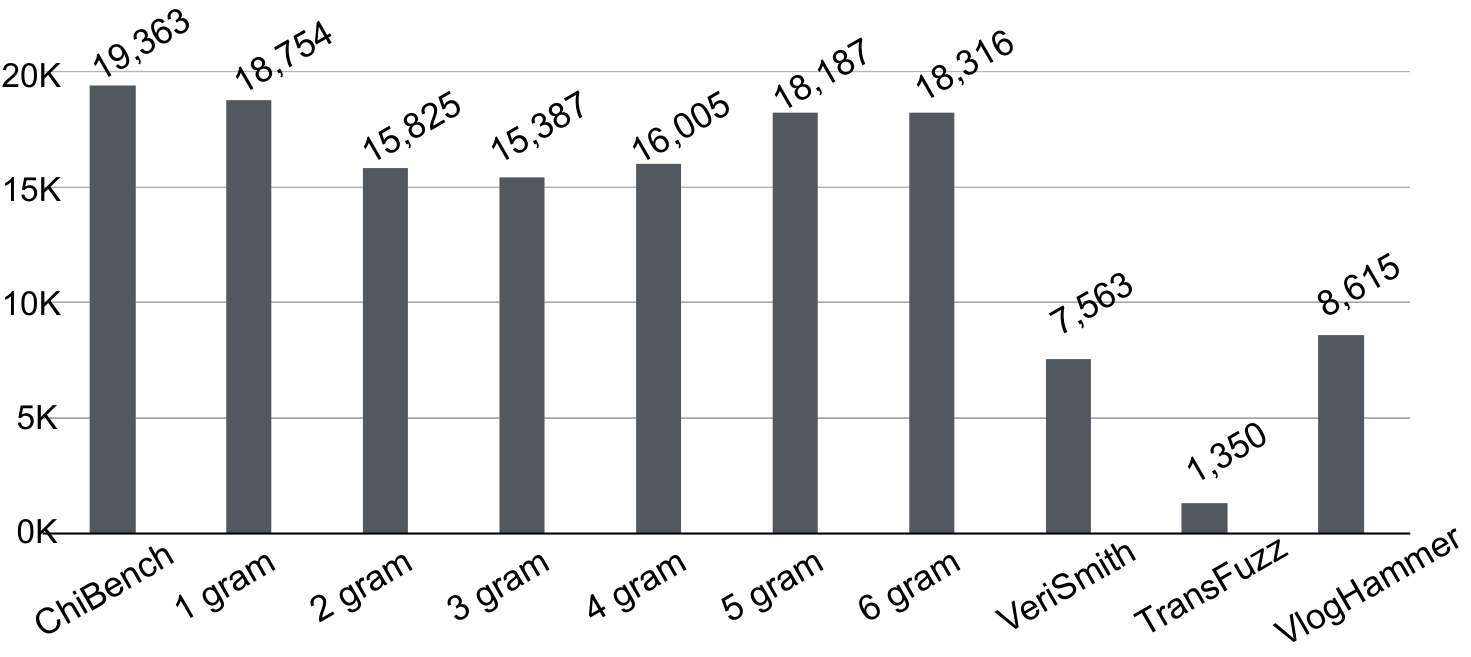}
\caption{Diversity of ChiGen designs, measured as the number unique 4-grams: the number of different sequences formed by four kinds of tokens.}
\label{fig_NGrams}
\end{figure}

Because ChiGen exercises more production rules than the other fuzzers, the programs it produces tend to be more diverse in terms of the number and type of tokens they use.
To demonstrate this fact, Figure~\ref{fig_NGrams} shows the number of 4-grams found within a population of 10,000 designs.
A 4-gram, in Figure~\ref{fig_NGrams} is a sequence of four kinds of tokens, e.g., ``$\mathit{ID} \rightarrow \mathtt{less\_than}\rightarrow \mathit{int}\rightarrow \mathtt{semi\_colon} $''.
The six populations produced by ChiGen, with six different probabilistic contexts, approach the diversity observed in ChiBench, which is formed by actual Verilog codes.
In contrast, the populations produced by the other fuzzers contain a very small number of different 4-grams.

%427 - Chibench
%137 - Vloghammer
%179 - Verismith
%456 - Verilog grammar

%1-gram productions (for 2^10, 2^11, 2^12, 2^13 and 2^14) 
%[331, 350, 376, 394, 406]
%2-gram productions
%[306, 338, 342, 366, 390]
%3-gram productions
%[313, 332, 340, 369, 391]
%4-gram productions
%[352, 354, 361, 379, 389]
%5-gram productions
%[354, 359, 368, 391, 400]
%6-gram productions
%[320, 323, 350, 385, 401]

\subsection{RQ2 -- Coverage}
\label{sub_coverage}

Code coverage in software testing measures the percentage of code executed during testing.
A test set is considered better than another if it enables higher coverage.
This section evaluates the coverage of ChiGen designs using two metrics: {\it branch coverage} and {\it line coverage}.
Branch coverage measures how many of the possible branches or decision points in a program have been executed by a set of test cases. In the context of conditional statements like \texttt{if} or \texttt{switch}, branch coverage checks whether each possible path (or ``branch''), such as both the true and false paths of an if statement, has been tested.
A coverage of 100\% indicates that every instruction of the binary program was fetched at least once during the execution of the test case.
Line coverage counts how many lines in the source code of the tested program were executed during the test.
This metric tracks the execution of executable lines in code, meaning it focuses on lines that contain actual instructions that are executed during runtime. Comments or blank lines do not count toward coverage.
Both metrics are measured via Clang's source-based code coverage feature (available in Clang 14.0.0).

\begin{figure*}[t!]
\centering
\includegraphics[width=1.0\textwidth]{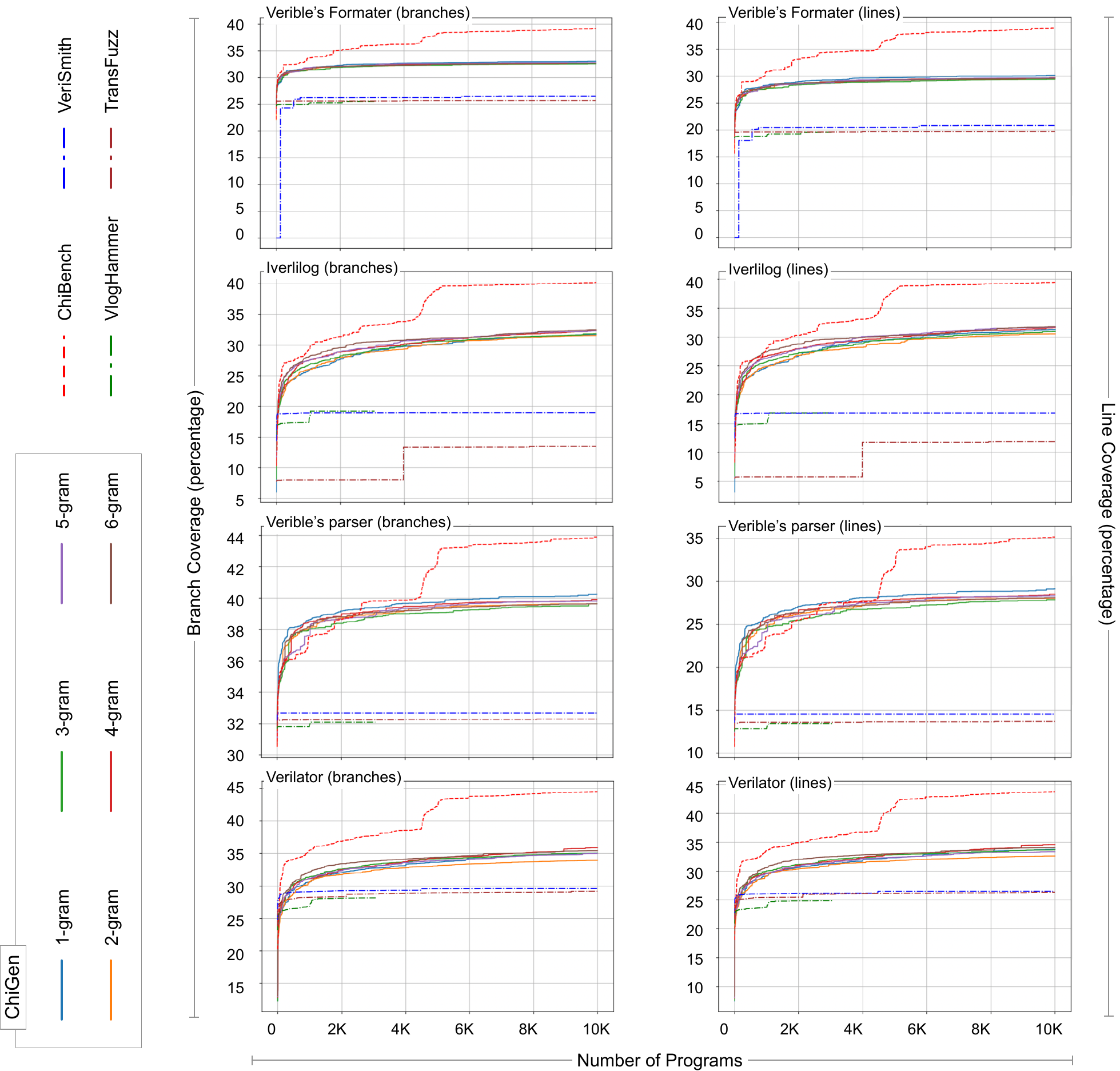}
\caption{Branch coverage obtained by testing Verible's syntactic analyzer with different sets of designs.}
\label{fig_coverage}
\end{figure*}

\paragraph{Discussion}
In each of the eight charts in Figure~\ref{fig_coverage}, the same trend is observed: ChiBench achieves the highest coverage, followed by ChiGen, VeriSmith, TransFuzz, and VlogHammer, in that order. The difference between ChiGen and the other fuzzers is noticeable; in some cases, such as Verible's parser, it results in nearly twice the coverage.

Another noteworthy observation is the consistent and gradual improvement in coverage with ChiGen designs. Indeed, coverage has not stabilized in any of the charts, although it reaches a plateau of small gains fairly quickly in Verible's formatter. However, this trend is not observed with the other fuzzers, which all seem to converge to a fixed percentage of code coverage after generating around 6,000 samples.
This observation corroborates the findings in Section~\ref{sub_diversity}, which suggest that ChiGen designs are more diverse than those produced by the other fuzzers.

The coverage achieved through human-made ChiBench designs outperforms that of synthetic benchmarks in every experiment. One reason for this superior performance is the SystemVerilog syntax. Currently, ChiGen's support for SystemVerilog is limited: it does not generate features such as classes and objects, fork-join, wait fork, dynamic and associative arrays, constrained random stimulus generation, clocking blocks, or interface types, for example.
This limitation is not dependent on the context of probabilities used in the probabilistic grammar: these features were intentionally omitted.
Incorporating them into ChiGen is more a matter of development priorities than an inherent theoretical challenge. Over time, additional SystemVerilog features will be integrated into the tool, and we hope that the open-source community will contribute to expanding ChiGen's capabilities in this area.

\subsection{RQ3 -- Bugs}
\label{sub_bugs}

ChiGen was designed as a qualification tool for the Jasper Formal Verification Platform and has been used for this purpose within Cadence Design Systems' development methodology.
The effectiveness of ChiGen in this context is classified information.
Nevertheless, ChiGen has been used to test a number of open-source EDA tools.
These tests have revealed issues, many of which were reported and acknowledged in public forums.
Table~\ref{tab_verilog_issues} lists some of these issues our group is aware of.

\begin{table*}[h]
    \centering
    \caption{Issues discovered in various verilog tools via tests using ChiGen designs.}
    \begin{tabular}{|c|l|l|}
        \hline
        \textbf{Issue} & \textbf{Tool} & \textbf{Description} \\ 
        \hline
        \href{https://github.com/chipsalliance/verible/issues/2159}{2159} & \href{https://github.com/chipsalliance/verible/tree/master/verible/verilog/tools/obfuscator}{Verible's Obfuscator} & The tool crashes when reading a design that only contains the pragma directive. \\
        \hline
        \href{https://github.com/chipsalliance/verible/issues/2181}{2181} & \href{https://github.com/chipsalliance/verible/blob/master/verible/verilog/parser/verilog.y}{Verible's Parser} & The tool crashes instead of reporting syntax errors related to instantiation type. \\
        \hline
        \href{https://github.com/chipsalliance/verible/issues/2189}{2189} & \href{https://github.com/chipsalliance/verible/blob/master/verible/verilog/tools/formatter/README.md}{Verible's Code Formatter} & The tool crashes with syntactically valid input. \\
        \hline
        \href{https://github.com/chipsalliance/verible/issues/2233}{2233} & \href{https://github.com/chipsalliance/verible/blob/master/verible/verilog/parser/verilog.y}{Verible's Parser} & The tool incorrectly accepts Verilog code with mismatched program and endmodule keywords. \\
        \hline
        \href{https://github.com/verilator/verilator/issues/5276}{5276} & \href{https://github.com/verilator/verilator}{Verilator} & The tool crashes with signal 9 on a very large design. \\
        \hline
        \href{https://github.com/verilator/verilator/issues/5311}{5311} & \href{https://github.com/verilator/verilator}{Verilator} & The tool crashes when using time assignments. \\
        \hline
        \href{https://github.com/verilator/verilator/issues/5312}{5312} & \href{https://github.com/verilator/verilator}{Verilator} & The tool crashes when calling a function created in a "generate" block. \\
        \hline
        \href{https://github.com/verilator/verilator/issues/5865}{5865} & \href{https://github.com/verilator/verilator}{Verilator} & The tool crashes when passing inout ports to primitive gates. \\
        \hline
        \href{https://github.com/steveicarus/iverilog/issues/1174}{1174} & \href{https://github.com/steveicarus/iverilog}{Icarus Verilog} & The tool crashes when assigning to parameters in a procedural block. \\
        \hline

        \href{https://github.com/steveicarus/iverilog/issues/1225}{1225} & \href{https://github.com/steveicarus/iverilog}{Icarus Verilog} & The tool freezes when computing invalid infinite loop.\\
        \hline
        
        \href{https://github.com/YosysHQ/yosys/issues/4598}{4598} & \href{https://github.com/YosysHQ/yosys}{Yosys} & The tool crashes while simplifying design. \\
        \hline
        \href{https://github.com/chipsalliance/verible/issues/2359}{2359} & \href{https://github.com/chipsalliance/verible/tree/master/verible/verilog/tools/formatter/README.md}{Verible's Code Formatter} & The tool fails to parse dash in front of unary operation (\texttt{- -1}). \\
        \hline
        \href{https://github.com/chipsalliance/verible/issues/2364}{2364} & \href{https://github.com/chipsalliance/verible/tree/master/verible/verilog/tools/formatter/README.md}{Verible's Code Formatter} & Fails to parse parameter declaration without qualifier (\verb+#(id_23=1)+). \\
        \hline
    \end{tabular}
    \label{tab_verilog_issues}
\end{table*}

To complement Table~\ref{tab_verilog_issues}, this section demonstrates how ChiGen compares to other fuzzers in terms of its ability to reveal crashes in EDA tools.
To this end, we have carried out bug-finding campaigns on different open-source tools: Verible (v0.0-3808)'s obfuscator; Yosys v0.45, and Verilator Release 159. In this process, we had to compile each tool with and without AddressSanitizer~(Asan)~\cite{Serebryany12}.
Each campaign consists on producing a population of 3,000 designs -- 500 for each production context using ChiGen -- and submitting these designs to each EDA tool. We classify as an ``issue'' any situation where a random Verilog design causes either a segmentation fault or a failed assertion.

\paragraph{Discussion}
Table~\ref{tab_errors} summarizes our findings.  When compiled with AddressSanitizer, ChiGen revealed 754 crashes in Yosys, compared to 747 from both Verismith and VlogHammer and 742 from TransFuzz; 52 against a single crash from both tools in Verilator and no crash from TransFuzz; and 766 in Verible's obfuscator where the peak was 771 by Verismith. Without Asan, ChiGen uncovered 47 crashes in Verilator, while Verismith and VlogHammer each found one. Finally, in Verible's Obfuscator, ChiGen detected 719 crashes, closely trailing Verismith's 728 and VlogHammer's 721.
Notice that we do not distinguish different crashes caused by the same issue, because that would require manually inspecting thousands of logs produced by AddressSanitizer.

% \usepackage{multirow}
\begin{table}[htb]
\caption{Crashes observed with 3,000 designs.
`Asan/!Asan': tool compiled with/without AddressSanitizer; CB: ChiBench; VS: Verismith; VH: VlogHammer; CG: ChiGen; TF: TransFuzz.}
\centering
\begin{tabular}{c|l|c|c|c|c|c|}
\cline{2-7}
\multicolumn{1}{l|}{}                        &           & \multicolumn{1}{l|}{CB} & \multicolumn{1}{l|}{VS} & \multicolumn{1}{l|}{VH} & \multicolumn{1}{l|}{CG} & \multicolumn{1}{l|}{TF} \\ \hline
\multicolumn{1}{|c|}{\multirow{3}{*}{Asan}}  & Yosys     & 712                           & 747                            & 747                             & 754                         & 742                            \\ \cline{2-7} 
\multicolumn{1}{|c|}{}                       & Verilator & 0                             & 1                              & 1                               & 52                          & 0                              \\ \cline{2-7} 
\multicolumn{1}{|c|}{}                       & Obfuscate & 734                           & 771                            & 728                             & 766                         & 761                            \\ \hline
\multicolumn{1}{|c|}{\multirow{3}{*}{!Asan}} & Yosys     & 0                             & 0                              & 0                               & 0                           & 0                              \\ \cline{2-7} 
\multicolumn{1}{|c|}{}                       & Verilator & 0                             & 1                              & 1                               & 47                          & 0                              \\ \cline{2-7} 
\multicolumn{1}{|c|}{}                       & Obfuscate & 0                             & 728                            & 721                             & 719                         & 0                              \\ \hline
\end{tabular}
\label{tab_errors}
\end{table}

\subsection{RQ4 -- Throughput}
\label{sub_throughput}

Throughput refers to the rate at which ChiGen produces test cases over a specific period of time.
The higher its throughput, the more designs ChiGen produces.
This section reports ChiGen's throughput as the time necessary to generate 100 designs, using different contexts of probabilistic productions.
We set the lower limit for the number of tokens to 150.

\begin{figure}[ht]
\includegraphics[width=\columnwidth]{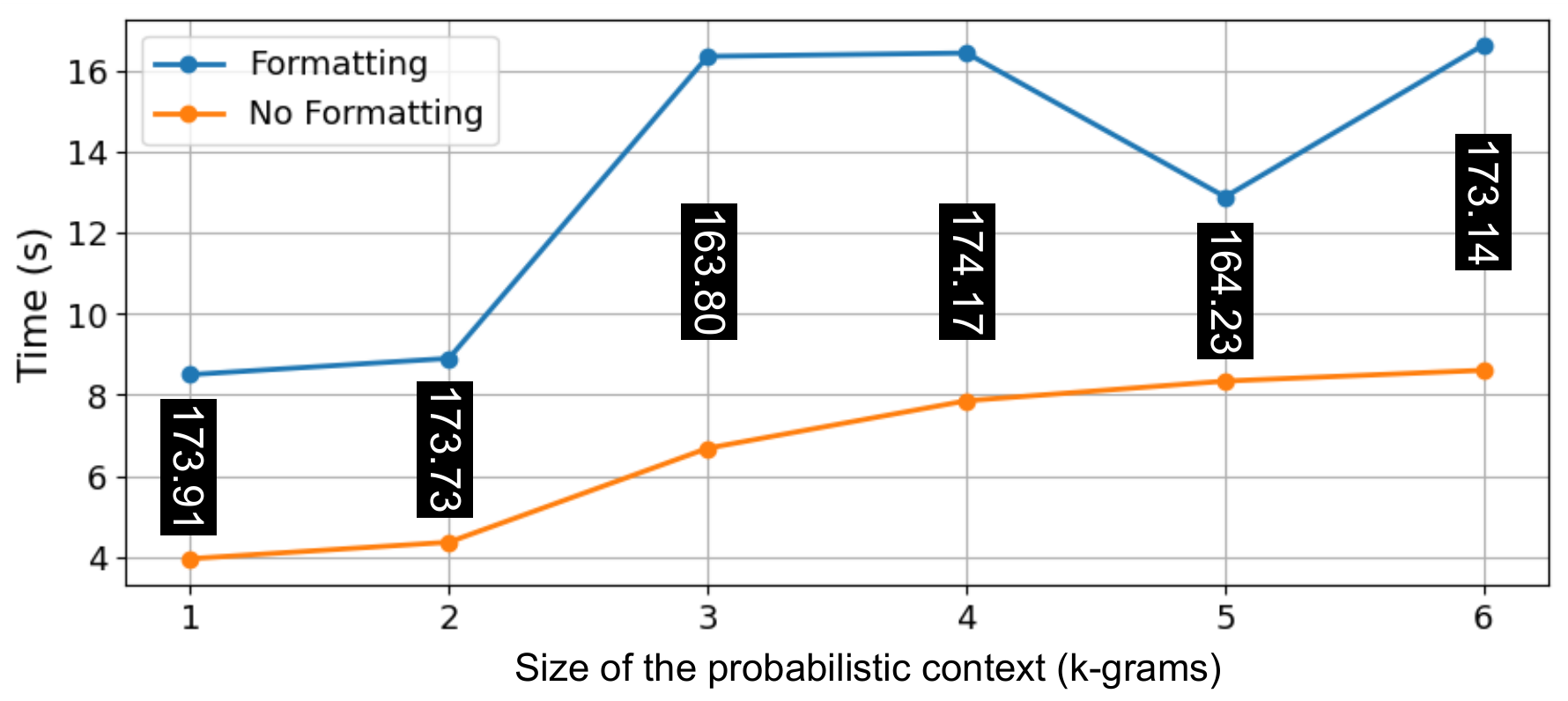}
\caption{Time to generate 100 designs, setting the lower limit of tokens to 150. Numbers in boxes show the average number of tokens.}
\label{fig_throughput}
\end{figure}

\paragraph{Discussion}
Figure~\ref{fig_throughput} illustrates the throughput of ChiGen, showing the time in seconds required to produce 100 designs across six sizes of probabilistic contexts.
The blue line represents the time elapsed when output codes are formatted using Verible’s formatting tool. We observe a noticeable overhead introduced by formatting, which peaks at contexts of size 3, 4, and 6 grams (in the 5-gram setting, we obtained fewer tokens on average, which reduces formatting time).
While formatting adds processing time, it enhances the readability of the generated code for manual analysis.
In contrast, the orange line shows the time required without formatting.
This line provides a more realistic picture of ChiGen's baseline performance.
Without formatting, the tool demonstrates a steady, gradual increase in processing time from 1-gram to 6-gram.
Nevertheless, even on its slowest configuration -- contexts of size 6 -- ChiGen can still produce about 100 valid Verilog designs with at least 150 tokens each in less than nine seconds on a commodity machine.

\subsection{RQ5 -- Software Evolution}
\label{sub_synthesis}

\begin{figure}[ht]
\includegraphics[width=1\columnwidth]{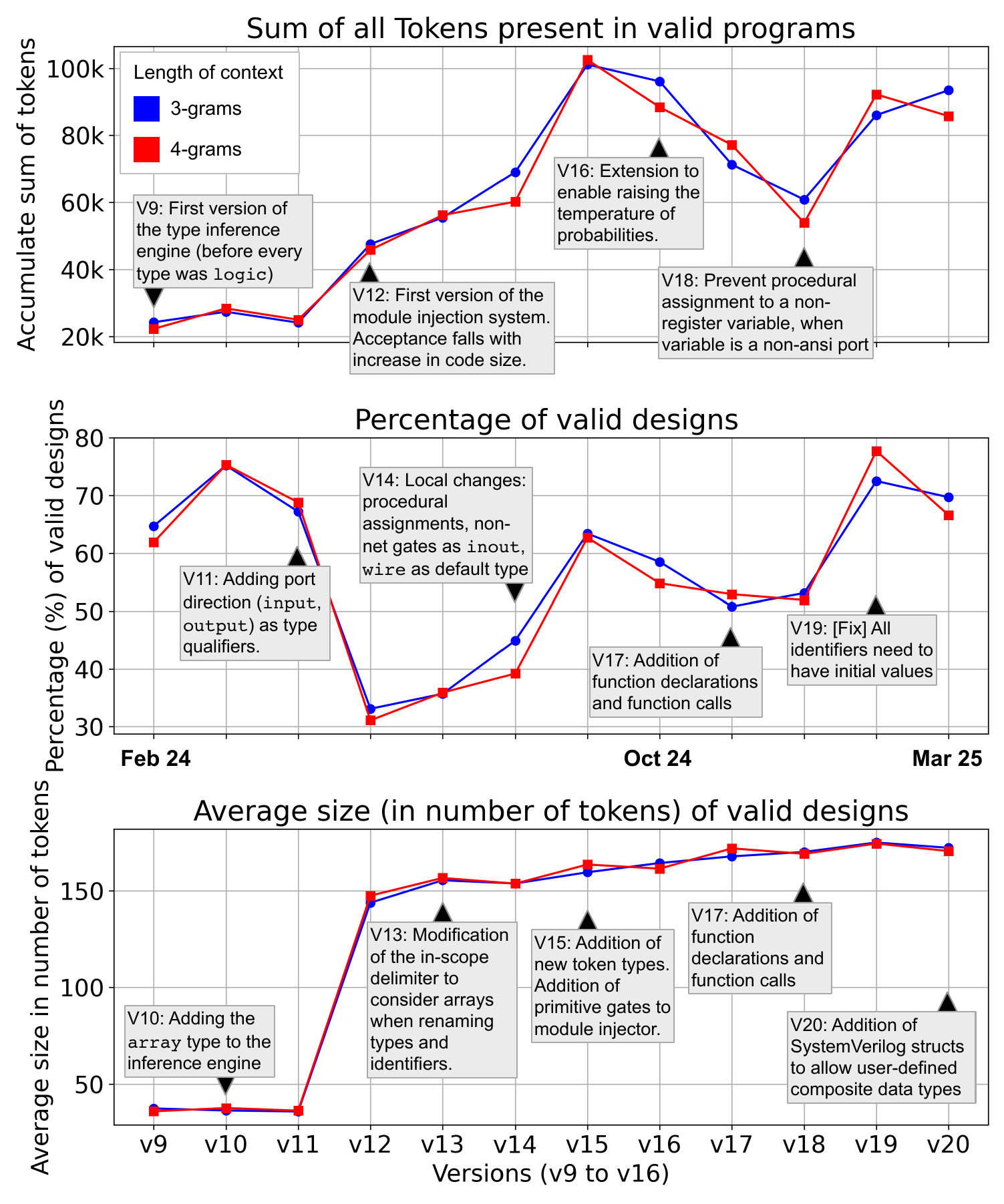}
\caption{The evolution of ChiGen, until its announcement as an open-source tool in October 2024. Each dot refers to 1,000 designs. Starting in v0.12, module injection uses a lower limit of 100 tokens.}
\label{fig_synthesis}
\end{figure}

ChiGen v0.09 introduced the type inference engine described in Section~\ref{sub_type_infer}, and since then, we have been tracking its effectiveness in generating valid Verilog designs.
When ChiGen was announced as an open-source tool in October 2024, it was at version 0.16. Figure~\ref{fig_synthesis} illustrates the evolution of ChiGen's capabilities over time.
The figure considers a population of 1,000 designs, where valid designs are those that pass the semantic analysis performed by Jasper.

Key milestones in this evolution include: (i) Completion of the type inference engine in v0.11; (ii) First release of the module injection engine (Section~\ref{sub_code_injection}) in v0.12; and (iii) Addition of an extension for varying production probabilities in v0.16. This last extension, not detailed in Section~\ref{sec_sol}, enables users to ``raise the temperature'' of probabilities -- effectively increasing the likelihood of triggering low-probability productions at the cost of slightly reducing the chances of higher-probability ones. This modification decreases the number of valid Verilog designs generated, but boosts their diversity.

\begin{figure}[ht]
\includegraphics[width=1\columnwidth]{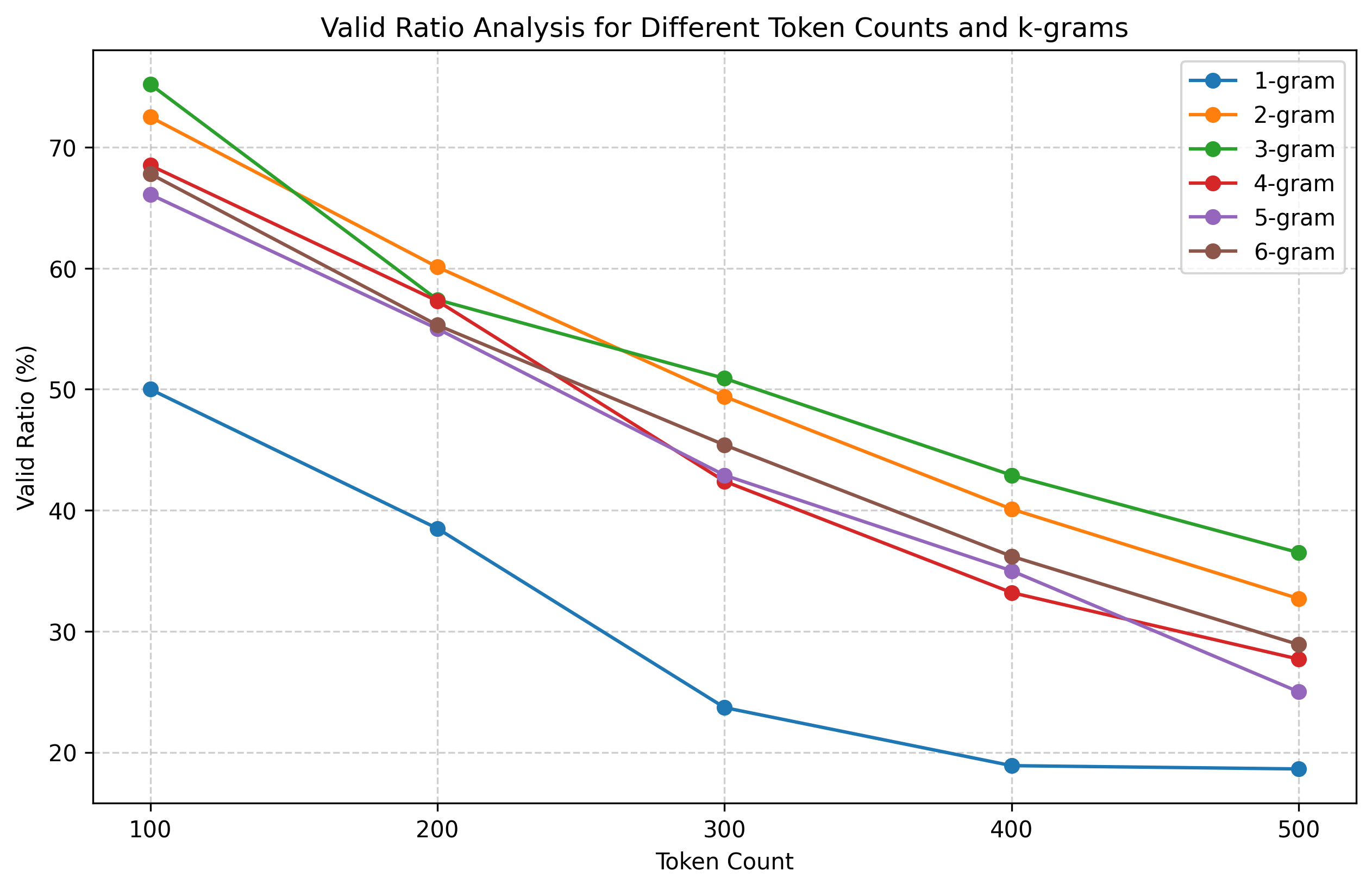}
\caption{Variation of the percentage of valid designs with the lower bound on the number of tokens in the designs produced by ChiGen.}
\label{fig_tokens_experiment}
\end{figure}

Code injection (v0.12) allowed us to increase substantially the size of designs that ChiGen produces; however, it also reduced the percentage of valid codes.
This reduction happened due to the probabilistic nature of the codes that ChiGen produces.
Thus, the larger the text it outputs, the higher the chance that invalid specifications will emerge.
Figure~\ref{fig_tokens_experiment} shows this trend.
According to the figure, if we set the lower limit of the number of tokens in 500, then we have about 30-40\% percent of valid designs with a probabilistic context of length two or three.
The proportion is much lower with a probabilistic context of length one, as the lack of context removes much of the syntactic constraints of the Verilog grammar.
Nevertheless, since v0.16 ChiGen has experienced constant evolution, and presently, the percentage of valid designs it produces is higher than pre-module injection (v0.11).

\section{Related Work}
\label{sec_rw}

\textbf{Fuzzers: }
This paper presents techniques for building Verilog fuzzers. Several other Verilog fuzzers are available as open-source tools~\cite{Herklotz20, Solt25, Wolf24}. In contrast to our work, these tools operate by gradually expanding a core set of Verilog syntax, ensuring that each expansion results in a valid design. During the development of ChiGen, we had the opportunity to engage with the authors of Verismith. We believe the following statement, shared in personal communication, highlights key differences between the two approaches: ``{\it I think the main difference between the two approaches, [$\ldots$] is that Verismith generates a set of Verilog modules one line at a time, making decisions locally about what statement to generate next. We also perform the entire generation in a single step, so splitting up different phases into separate steps is very interesting. Again, this could be seen as a local approach versus a more global and modular approach in ChiGen.}~\cite{Herklotz24}''

\textbf{LLMs: }
Over the last two years, the LLM revolution has brought to light many language models for Verilog~\cite{Gao24, Liu23, thakur_verigen_2024,Wang24}. ChiGen is not a language model; it is a fuzzer, meaning that it does not attempt to shape the code toward any specific semantics. ChiGen does use a probabilistic grammar that models the probability of production rules as k-grams; however, it does not assign probabilities to sequences of tokens -- rather, it assign them to sequences of production rules.

\textbf{Benchmarks: }
ChiGen can be used to generate Verilog benchmarks; however, it is not a benchmark collection.
There already exist collections of benchmarks formed by hardware specification languages~\cite{Amaru19, Babb97, Brglez89, Kozminski91, Murray15, Sumitani24}.
In contrast to ChiGen, these collections are {\it immutable}: these previous works do not generate new benchmarks and incorporate them automatically into the database of available codes.

\section{Conclusion}
\label{sec_conc}

This paper introduced the design of what we call a "bottom-up" fuzzer: a tool that generates Verilog specifications by first constructing a skeleton of Verilog syntax and then completing this skeleton by inferring names and types.
While these techniques were applied in the context of Verilog, we believe that the combination of probabilistic grammars, Hindley-Milner type inference, and Li-Zhendong code injection can automate code generation for any programming language.
Although none of these techniques represents a novel contribution, their synergistic integration as a method for implementing Verilog design fuzzers is unique.

It has always been an internal goal of this project that about 60-70\% of every Verilog specification produced by ChiGen should be valid: the rest would have either syntactic or semantic faults -- this decision has guided much of ChiGen's design.
As discussed in Section~\ref{sec_ovf}, the invalid outputs generated by ChiGen have uncovered zero-day bugs in several EDA tools. For instance, during a discussion on an issue ChiGen uncovered in Icarus, one of the developers commented: ``{\it When you think about the psychology of code development, this likely makes sense. Sure, there are checks for certain invalid cases, but we can often make assumptions that our users will not get too far from valid code.}~\cite{Cary24}''

ChiGen generates some SystemVerilog syntax; however, it does not support the full IEEE 1800-2017 SystemVerilog specification.
This limitation is a deliberate design choice by ChiGen's developers, who have chosen to focus exclusively on the IEEE Standard for Verilog Hardware Description Language (Verilog-2005)~\cite{IEEE06}.
Nevertheless, the underlying principles of ChiGen could be extended to support other languages, including SystemVerilog, as well as general-purpose programming languages like C or Java. Expanding ChiGen's capabilities remains a possibility for future development, depending on evolving priorities and contributions.

\bibliographystyle{plain}
\bibliography{references}

\end{document}